\documentclass[aps,prl,twocolumn,showpacs,preprintnumbers]{revtex4-1} 
\usepackage{graphicx,epsfig}
\usepackage{bm}
\usepackage{latexsym,amssymb,amsmath,float}
\usepackage{subfig}
\usepackage[countmax]{subfloat}
\usepackage{color}
\newcommand{\beq}[1]{\begin{equation}\label{#1}}
\newcommand{\eeq}{\end{equation}}
\newcommand{\bea}{\begin{eqnarray}}
\newcommand{\eea}{\end{eqnarray}}
\newcommand{\ba}{\begin{array}}
\newcommand{\ea}{\end{array}}

\newcommand{\rf}[1]{(\ref{#1})}
\def\be{\begin{equation}}
\def\ee{\end{equation}}
\def\gs{\mathrel{
   \rlap{\raise 0.511ex \hbox{$>$}}{\lower 0.511ex \hbox{$\sim$}}}}
\def\ls{\mathrel{
   \rlap{\raise 0.511ex \hbox{$<$}}{\lower 0.511ex \hbox{$\sim$}}}}

\newcommand{\lsim}{\lesssim}
\newcommand{\unitmatrix}{\openone}

\newcommand{\half}{\frac{1}{2}}

\newcommand{\bad}{\begin{array}{ccc}}
\newcommand{\bav}{\begin{array}{cccc}}
\newcommand{\baf}{\begin{array}{ccccc}}

\newcommand{\nue}{\nu_{e}}

\newcommand{\Upmns}{U_{\rm PMNS}}

\def\Det{{\rm Det}}

\renewcommand{\tilde}{\widetilde}

\def\T{{\sf T}}

\def\S{{\sf S}}

\begin{document}


\title{Simple Mass Matrix Ans\"atze for Neutrino Mixing}

\author{HoSeong La}
\author{Thomas J. Weiler}
\affiliation{Department of Physics and Astronomy,
Vanderbilt University, Nashville, TN 37235, USA \\
{\tt hsla.avt@gmail.com} \hskip16pt {\tt tom.weiler@vanderbilt.edu}
}

\date{November 6, 2013}

\begin{abstract}
We estimate the PMNS (Pontecorvo-Maki-Nakagawa-Sakata) matrix in terms 
of neutrino and charged lepton mixing given as
$U_{\rm PMNS}=V_\ell^\dagger(\tilde\theta_{13}) U_\nu(\tilde\theta_{23}, 
\tilde\theta_{12})$, based on a new (emergent) global lepton flavor symmetry. 
The neutrino and charged lepton mass matrices have simple textures.
The resulting $U_{\rm PMNS}$ gives excellent agreement with experimental
data (including $|U_{e3}|\simeq 0.16$).
\end{abstract}


\maketitle

\setcounter{footnote}{0}


The Standard Model (SM) of the Electroweak (EW) and 
Quantum Chromodynamics (QCD) has been amazingly successful. 
Despite the successes, it is well known that the SM 
is not an ultimate fundamental theory. 
The presence of dark matter and dark energy in the Universe,
and neutrino flavor oscillations\cite{pdg} pose challenges to the SM. 
The latter are commonly interpreted to indicate that at least two
neutrinos are massive and all three masses are non-degenerate;
the flavor-violations result from the different relation between the 
flavor and mass eigenstates of the neutrinos relative to the relation for the charged leptons\cite{Pontecorvo:1957qd}.
It is thus important to understand the structure of these neutrino masses and mixings, as it offers a portal to physics beyond the SM.
In this paper we will present simple mass ans\"atze which incorporate
the observed data 
(including nonzero $\theta_{13}\sim 9^\circ$)\cite{pdg2013}. 
Then, we present a new (emergent) symmetry based on a product of cyclic groups 
which enforces the mass ans\"atze.

The PMNS 
mixing matrix in the lepton sector 
is a result of the mismatch between the left-multiplying matrices 
which diagonalize the charged lepton mass matrix ($V_\ell$) 
and the neutrino mass matrix ($U_\nu$),
\beq{UisVU}
\Upmns=V_\ell^\dag U_{\nu}\,.
\eeq
Under these transformations of bases, the charged current (Weak) 
interactions become
\begin{equation}
\label{CC}
{\cal L}_{\rm CC}= \tfrac{g_2}{2} W_\mu \overline{V_\ell}
\overline{\ell} \gamma^\mu P_L U_\nu \nu
= \tfrac{g_2}{2} W_\mu \overline{\ell}\gamma^\mu 
\Upmns P_L\, \nu,
\end{equation}
where $P_L\equiv\half(1-\gamma_5)$ is the left-handed chiral projector.
As is often said, $\nu_\alpha = (\Upmns)^*_{\alpha j} \nu_j$ 
is the neutrino state of flavor~$\alpha$ 
in the basis where the charged lepton mass matrix is diagonal.

Later in this paper, we will need to accommodate our result to the standard 
PDG form\cite{pdg}, since that ordering defines the mixing angles adopted 
by the experimental community.
So we present the PDG form here:
%
\beq{PDG1} 
\Upmns=R_{23}(\theta_{23})\,U_\delta^\dag\,R_{13}(\theta_{13})\,U_\delta\,R_{12}(\theta_{12}) \,,
\eeq
where $U_\delta$ a $CP$-violating phase matrix.
In the Dirac case,
there results
\bea
\label{PDG2}
\Upmns &=& \left(
\ba{ccc}
c_{12} c_{13} & s_{12} c_{13} & 0 \\
-s_{12} c_{23} & c_{12} c_{23} & s_{23} c_{13} \\
s_{12} s_{23} & -c_{12} s_{23} & c_{23} c_{13} 
\ea
\right) \nonumber \\
&-& \left(s_{13}\,e^{+i\delta} \right) 
\left(
\ba{ccc}
 0 			    & 0 			     & -e^{-2i\delta} \\
 c_{12} s_{23} & s_{12} s_{23} & 0 			    \\
 c_{12} c_{23} & s_{12} c_{23} & 0
 \ea
 \right)
\eea
in obvious notation\footnote{It can be useful to separate $\Upmns$ 
as we do to illustrate that 
(i) the phase always enter with $s_{13}$ a prefactor;
(ii) since $\theta_{13}$ is known to be small, $\sim 0.16$, 
$\theta_{13}$ enters the first matrix first at second order, 
$\theta_{13}^2\sim 0.024$, but enters the second matrix at first order.
}.
By convention, the three  neutrino mass-eigenstates are labeled in reverse 
order to their $\nue$ content:
$|U_{e1}|^2 > |U_{e2}|^2 > |U_{e3}|^2$, where the $U_{ej}$ are the first-row 
matrix elements of $\Upmns$.
This implies that $c_{12}^2> s_{12}^2 > t_{13}^2$,
hence $|\theta_{13}|<\theta_{12}<\pi/4$,
while the octant of $\theta_{23}$ is not presently determined.

For a general charged leptons mass matrix $\mathbf{M}_\ell$ 
(not necessarily hermitian), 
one requires two independent matrices, $V_L$ and $V_R$, for diagonalization: 
viz., $V_L^\dag\mathbf{M}_\ell V_R = {\rm diag}(M_e, M_\mu, M_\tau)$.
Similar considerations apply to the neutrino mass matrix if the neutrinos are 
Dirac particles; if the neutrinos are Majorana, then 
$U_{\nu R} = U_{\nu L}$ and diagonalization occurs via a single matrix.
Only if Dirac mass matrices are themselves ``normal'', meaning that 
$[\mathbf{M},\mathbf{M}^\dag]=0$, 
are they diagonalizable by a similarity transformation with the 
single matrices $V_L\equiv V_\ell$ and $U_{\nu L}\equiv U_\nu$, respectively.
Note that hermitian matrices provide trivial examples of normal 
matrices.

An estimate of the number of parameters needed in the charged lepton and 
neutrino flavor-basis mass matrices to yield the three observed 
charged lepton masses, 
the three neutrino mixing angles, and the two mass-squared differences 
inferred from neutrino oscillations is eight.
In this paper we show that just seven parameters and a non-PDG ordering of planar rotations does reproduce the eight observables 
when the leptonic mass matrices in the Weak eigenstate basis
are real-symmetric with the forms\cite{La:2013gga}:
%
\begin{subequations}
\begin{align}
\label{Mlep}
\!\!\!\!\!\!
\mathbf{M}_\ell \! &= \!\!
\left(
\ba{ccc}
M_{11} & 0 & M_{13} \\
0           & M_\mu & 0 \\
M_{13} & 0 & M_{33} \\
\ea
\right), \\
\label{Mnu}
\mathbf{m}
\! &= \!\!
 \left(
\ba{ccc}
{m}_{11} & {m}_{12} & {m}_{13} \\
{m}_{12} & {m}_{11} & 0 \\
{m}_{13} & 0 & {m}_{11} \\
\ea
\right) \!\!
=\! {m}_{11}\unitmatrix \! +\! {m}_{12} \!
\left( 
\ba{ccc}
0 & 1 & r \\
1 & 0 & 0 \\
r & 0 & 0 \\ 
\ea
\right)\! ,
\end{align}
\end{subequations}
where $M_{22}$ is equal to the physical muon mass $M_\mu$,
$\mathbf{m}$ generically denotes the neutrino mass matrix relevant 
to the observations, i.e.
$\mathbf{m}_{\rm D}=\mathbf{m}$ if physical neutrinos are Dirac 
while $\mathbf{m}_L=\mathbf{m}$ if Majorana, 
and
we have defined $r\equiv {{m}_{13}}/{{m}_{12}}$. 
These textures of charged lepton and neutrino mass matrices
can be fixed upon us by symmetry considerations, as we will show. 


The ability of describing all eight inferred values with just 
seven parameters can be traced back to the extra symmetry implicit 
in the common diagonal of the neutrino mass matrix; this common diagonal 
fixes one of our mixing angles $\tilde{\theta}_{12}$
(to the maximum value of $45^\circ$), and yet leaves the parameter available 
for fixing another observable.
In this sense, our seven parameter matrices can be viewed 
as eight-parameter matrices with two parameters identified
via an internal symmetry.

Two remarks should be made at this point.
First, even though the value of one of our mixing angles is 
fixed to be maximal, $\tilde{\theta}_{12}=45^\circ$, 
no resulting angle in the PDG ordering will generally be equal to $45^\circ$.
Second, since in this paper we assume real symmetric mass 
matrices, there is no (complex-valued) $CP$-violation in our mixing matrix.
A phase could be added to either (hermitian) matrix to accommodate 
$CP$-violation, 
but since $CP$-violation in the lepton sector is phenomenologically 
unconstrained at present, such an addition seems premature.

We begin with diagonalization of the charged lepton mass matrix.
The three free parameters in the matrix are 
constrained by ${\rm Tr}\,\mathbf{M}_\ell=M_e +M_\mu +M_\tau$ and 
${\rm Det}\,\mathbf{M}_\ell = M_e M_\mu M_\tau$.
Here we take $M_{11}=M_\mu$ for definiteness (in addition to $M_{22}=M_\mu$),
and find that 
\be
\label{M13andM33}
\begin{aligned}
M^2_{13} &= (M_\tau-M_\mu)(M_\mu-M_e),\\
M_{33} &= M_\tau-M_\mu+M_e\,.
\end{aligned}
\ee
The choice $M_{11}=M_\mu$ is natural in the sense that it equates 
$M_{11}$ and $M_{22}$; versus, e.g., the choice $M_{11}=M_e$ 
which would have set $M_{13}=0$.

The matrix $\mathbf{M}_\ell$ in Eq.\rf{Mlep} is real-symmetric, 
and therefore trivially hermitian and normal; consequently, 
it is diagonalized by the single matrix $V_\ell$ which can be set 
equal to the rotation matrix $R_{13}(-\tilde\theta_{13})$ 
with angle given implicitly by%
\cite{La:2013gga}:  
\beq{tildetheta13}
\sin\tilde\theta_{13}=\sqrt{\frac{M_\mu-M_e}{M_\tau-M_e}}\,.
\eeq
%
Thus, we have $\Upmns=V^\dag_\ell U_\nu=R_{13}(\tilde\theta_{13})\,U_\nu$.
Next we turn to diagonalization of the neutrino mass matrix.

Currently, it is not known whether neutrinos are Majorana or Dirac in nature.
Majorana neutrinos are favored, because they generally result from 
the see-saw mechanism which justifies the small neutrino mass naturally. 
In fact, later
we will find that the Majorana case is also more favored by symmetry. 

Diagonalizing the mass matrix $\mathbf{m}$ yields the 
neutrino masses: 
\be
\label{e:36a}
m_1=m_-,\ m_2=m_+,\ m_3=m_{11}, 
\ee
where
$m_\pm\equiv m_{11}\pm\sqrt{m^2_{12}+m^2_{13}}$,
and that
\be
\label{e:36b}
2m_3=m_1 + m_2\,,\ {\rm and\ ordering\ }m_1  <  m_3  <  m_2 \,.
\ee
Note that the neutrino masses satisfy equal spacing,
but {\sl a priori}, one or more of them 
can be negative. 

The solar neutrino measurement\cite{SNO} that the ratio of 
the charged-current to neutral-current cross sections are $\sim 1/3$ leads, 
by virtue of matter effects in the Sun, to the inference that the heavier 
of $m^2_1$ and $m^2_2$ has less $\nue$ content, i.e., that $m_2^2 > m_1^2$.
This in turn fixes the sign of $m_{11}=m_3$ to be positive.

The ordering of $m^2_3$ relative to $m^2_2$ and $m^2_1$ 
is not known. However, in our 
model, with $m_3$ bounded from below 
by $m_1$ and from above by $m_2$, the equal-spacing of the masses cannot
meet the condition for a ``normal'' hierarchy, 
$m^2_3\gg \{m^2_2, m^2_1\}$.
We may however, have an ``inverted'' hierarchy, 
characterized by $m^2_3\ll \{m^2_2, m^2_1\}$.

%
%

The diagonalization of 
$\mathbf{{m}}$ 
given in Eq.\rf{Mnu} is effected by setting 
$U_\nu= R_{23}(\tilde\theta_{23})R_{12}(\tilde\theta_{12}) $, 
with angles given by%
\beq{tildetheta12}
\tan\tilde\theta_{23} =	-r\,, \quad{\rm and\ \ } 
\tilde\theta_{12} =  \tfrac{\pi}{4} \,.   
\eeq
We note that were ${m}_{23}^2$ taken to be equal to 
${m}_{12}^2$, i.e., $r^2=1$, then the resulting matrix is diagonalized 
by the bimaximal (BM) mixing matrix\cite{BPWW}.

Having found the diagonalizing matrices for the charged lepton and neutrino 
mass matrices, we arrive at the neutrino mixing matrix:
\bea
\label{Upmns1}
\Upmns &=& R_{13}(\tilde\theta_{13})\,R_{23}(\tilde\theta_{23})\,
R_{12}(\tilde\theta_{12}) \\
&=& R_{13}\left(\sim\sqrt{\tfrac{M_\mu}{M_\tau}}\,\right) 
R_{23}\left(-\tan^{-1}r\right)
R_{12}\left(\tfrac{\pi}{4}\right)  \,. \nonumber
\eea
Comparison with the PDG form in Eq.\rf{PDG2} leads to
\beq{translate2W}
t_{23} =\frac{\tilde{t}_{23}}{\tilde{c}_{13}}\,,\ \ 
t_{12} ={\tilde{t}_{12} -\tilde{t}_{13} \tilde{s}_{23} \over 
1+\tilde{t}_{13} \tilde{t}_{12} \tilde{s}_{23}}\,,\ \ 
s_{13} =\tilde{s}_{13} \tilde{c}_{23}\,.
\eeq
At this stage, $r$ is undetermined. Without any symmetry argument, we will 
have to use one observed angles as an input value to estimate the rest. 
For example, with the observed $\theta_{23}$, 
the first equation in eq.(\ref{translate2W}) leads to
%
\beq{r}
-r=\tilde{t}_{23} = t_{23} \tilde{c}_{13}\simeq 
\begin{cases}
1.22\ \ \ \, \mbox{for } \theta_{23}\simeq 51.5^\circ ,\\ 
0.797\ \ \mbox{for } \theta_{23}\simeq 38.5^\circ .
\end{cases}
\eeq
The two values given here for $\theta_{23}$ correspond to
$\sin^2(2\theta_{23})=0.95$; thus, $-r$ may lie between them.
See Table~\ref{table:one} for our results.

We make some comments on the change in the results here due to the nonzero 
value of $\theta_{13}$. First, were $\theta_{13}$ exactly
zero, then the ordering of our rotations coincides with that of the PDG;
consequently, one gets 
$\tilde\theta_{23}=\theta_{23}$ and 
$\tilde\theta_{12}=\theta_{12}$, which is not viable.
With $\theta_{13}$ now known to be far from zero, 
the ordering of our rotations is different from that of the PDG convention,
and our results are phenomenologically viable.
Second, our model assigns the nonzero $\tilde\theta_{13}$ rotation 
 entirely to diagonalization of the charged lepton matrix
(via $V_\ell=R(-\tilde\theta_{13}$)).
Third, in the present model, $\theta_{13}$ is naturally of order 
$\sqrt{{M_\mu/M_\tau}}$, 
a relationship noted in \cite{La:2013gga} for the first time.
Fourth, $r$ is exactly equal to 1 in the BM case\cite{BPWW} of 
$\theta_{13}=0$ and 
$\tilde\theta_{23}=\pi/4=\tilde\theta_{12}$, but deviates from 1 
in the presence of newer values for these two angles.

%
%

{\sl The Symmetry:}
We can justify  the forms of the mass matrices in 
Eqs.(\ref{Mlep})(\ref{Mnu}) in terms of a symmetry.
The symmetry 
is based on a finite group of a product of 
$C_2$'s, cyclic groups with order two.
$C_2(g)$ is the simplest possible nontrivial group because it contains only 
one extra group element $g$ such that $g^2=1$ in addition to the 
identity\cite{hamgp}, and yet it is sufficient for our purpose. 
To be more precise,
$C_2(\T')\times C_2(\T)\times C_2(\S)$ in the Dirac case, 
where $\T', \T$ and $\S$ are group generators for each $C_2$ that 
nontrivially act on the charged leptons, left-handed neutrinos, and
right-handed neutrinos, respectively, while
$C_2(\T')\times C_2(\S')$ in the Majorana case, where
$\S'$ acts on the left-handed Majorana neutrinos.

The mass matrices are invariant as (see \cite{La:2013rra} for more details)
\begin{subequations}
\begin{align}
\label{e:7a}
{\T'}^\dagger \mathbf{M}_\ell \T' &= \mathbf{M}_\ell, \\
\label{e:7b}
\T^\dagger \mathbf{m}_{\rm D} \S  &= \mathbf{m}_{\rm D} \ \ 
\mbox{in the Dirac case,}\\
\label{e:7c}
\S'^\dagger \mathbf{{m}}_L \S'^* &= \mathbf{{m}}_L \ \ 
\mbox{in the Majorana case,}
\end{align}
\end{subequations}
where ${\T'}^2=1$, ${\T}^2=1$, $\S^2=1$, and $\S'^2=1$. 
Note that $\T', \T$, $\S$ and $\S'$ 
must be also unitary so that these discrete 
symmetries can be global lepton flavor symmetries of the Lagrangian. 
Then they are also hermitian. 
Note that these discrete symmetries are unbroken and based on the low energy 
parameters related to the physical observables so that they should be regarded
as ``emergent symmetries"\cite{Chadha:1982qq,esmore}
for the U(1)$_{\rm em}$ lagrangian 
(albeit broken by Weak interactions only, similarly to the emergent parity
symmetry),
 but they are
not necessarily symmetries of the SM model before the EW symmetry breaking.
Imposing an unbroken emergent symmetry below the EW scale distinguishes 
our approach from others using discrete symmetries to constrain the 
neutrino masses and mixngs (see,
for example, the recent review \cite{King:2013eh} and references therein).

We proceed to establish the required forms of the generators in matrix representations.
For notational convenience, let us first define three basic operators:
\begin{align}
\Gamma_1(\theta_1)
&\equiv\left(
\begin{array}{ccc}
1 & 0 &0\\ 
0 &c_1  &s_1\\
0 &s_1 &-c_1
\end{array}
\right), \ 
\Gamma_2(\theta_2)
\equiv\left(
\begin{array}{ccc}
c_2 & 0 &s_2\\ 
0 &1  &0\\
s_2& 0 &-c_2
\end{array}
\right), \nonumber\\
\Gamma_3(\theta_3)
&\equiv\left(
\begin{array}{ccc}
c_3 &s_3& 0 \\ 
s_3 &-c_3 & 0\\
0  &0&1 
\end{array}
\right), \quad
{c_j\equiv\cos\theta_j,\atop s_j\equiv\sin\theta_j\ .}
\end{align}
$\Gamma_j$ is a combination of a rotation and an inversion 
and acts about $j$-axis, satisfying $\Gamma_j^2=1$, and $\Det\Gamma_j=-1$.

{\sl Dirac Case:}
The symmetry $C_2(\T')$ is generated by 
\begin{equation}
\label{e:8}
\T'=\Gamma_2(\theta_\ell),\quad \theta_\ell =2\tilde{\theta}_{13},
\end{equation}
where $\tilde{\theta}_{13}$ is given in eq.(\ref{tildetheta13}). 
Then eq.(\ref{e:7a}) fixes $M_{11}=M_\mu$ such that the last equation in 
eq.(\ref{translate2W}) relates $\tilde{\theta}_{13}$ to $\theta_{13}$.
This $\T'$ also ensures $M_{12}=0=M_{23}$,
so $C_2(\T')$ constrains $\mathbf{M}_\ell$ to be of 
the form given in eq.(\ref{Mlep})\cite{La:2013rra}.

For $C_2(\T)$ we choose
\begin{equation}
\label{e:8x}
\T=\Gamma_2(\theta_\nu ),\quad {\rm where\ } c_\nu  =\tilde{t}_{13}\, .
\end{equation}
And $C_2(\S)$ that matches $C_2(\T)$ is generated by 
\begin{subequations}
\begin{align}
\label{e:34u}
\S &=\Gamma_1(\theta_\alpha)\Gamma_3(\theta_\nu )\Gamma_1(\theta_\alpha)
\\
&=\left(
\begin{array}{ccc}
c_\nu  & s_\nu  c_\alpha &s_\nu  s_\alpha\\ 
s_\nu  c_\alpha &s_\alpha^2-c_\nu  c_\alpha^2& -c_\alpha s_\alpha (1+c_\nu )\\
s_\nu  s_\alpha &-c_\alpha s_\alpha (1+c_\nu ) &c_\alpha^2-c_\nu  s_\alpha^2
\end{array}
\right),
\end{align}
\end{subequations}
where
$s_\alpha = {1-3c_\nu \over 1+c_\nu }$, then eq.(\ref{e:7b}) leads to
\begin{equation}
\label{e:m1}
{m_{12}^2\over m_{11}^2}={8c_\nu \over 1+c_\nu },
\end{equation}
and, most importantly, an equality relating the mixing angle 
$\tilde{\theta}_{23}$
to the symmetry parameter
\begin{equation}
\label{e:18}
-r=\tilde{t}_{23}=\sqrt{1-c_\nu  \over 2 c_\nu }.
\end{equation}
Hence $r$ is no longer left undetermined.
This $\S$ fixes $\mathbf{m}_{\rm D}$ to be the form of eq.(\ref{Mnu}), 
which will lead to $\tilde{\theta}_{12}=\pi/4$. 

Eqs.(\ref{e:m1})(\ref{e:18}) imply that two of the Dirac neutrino mass-squared 
eigenvalues are degenerate such that $m_1=-m_{11}=-m_3$, $m_2\simeq 3m_{11}$.
As a neutrino mass degeneracy violates phenomenological findings,
Dirac type neutrinos are disfavored by the symmetry we impose.

{\sl Majorana Case:}
However, in the physical left-handed Majorana case, the 
masses are nondegenerate due to a different symmetry constraint
given by
eq.(\ref{e:7c}) with
\begin{equation} 
\label{e:sp}
\S'=\Gamma_1({\theta}_{\rm M})\Gamma_3(\tfrac{\pi}{2})
\Gamma_1({\theta}_{\rm M}), \ \ 
s_{\rm M}=-\sqrt{\tan(\tfrac{\pi}{4}-\tilde{\theta}_{13})}
\end{equation}
such that $\Gamma_3$ leads to $\tilde{\theta}_{12}=\pi/4$
while $\Gamma_1$ leads to eq.(\ref{e:18}).
In this case, there is no constraint like eq.(\ref{e:m1}) so that 
non-degenerate masses as eq.(\ref{e:36a}) can be obtained.

Having established the relative constraints between mixing angles, 
we can now estimate them.
Numerically, Eqs.\rf{e:8x}\rf{tildetheta13} fix the parameter 
$c_\nu  \simeq 0.2508$, or Eqs.(\ref{e:sp})\rf{tildetheta13} fix
$s_{\rm M}^2=0.5990$,
and thereby fix the symmetry.
As consequences, eq.(\ref{e:18}) leads to $\tilde{s}^2_{23}\simeq 0.599$,
i.e.  $\tilde{\theta}_{23}\simeq 50.7^\circ$,
hence fixing $-r\simeq 1.222$.
One of the beautiful features of the model is that 
$\tilde{\theta}_{13}$ is dominantly determined by the muon-to-tau mass ratio,
as given in Eq.\rf{tildetheta13}, such that
$\tilde{\theta}_{13}\simeq 0.243\simeq 14.1^\circ$,
and we arrive at 
\beq{Upmns2}
\Upmns = R_{13}(14.1^\circ) R_{23}(50.7^\circ) R_{12}(45^\circ) \,
\eeq
with our ordering of the three planar rotations.

Finally, we use the translations in Eq.(\ref{translate2W}) to obtain 
the PDG values 
$s^2_{12}\simeq 0.313$, $s^2_{23}\simeq 0.613$ and $s^2_{13}\simeq 0.0237$.
which are in excellent agreement with the observed data 
(see Table~\ref{table:one}).
The latter result follows from 
$\theta_{13} \approx s_{13}= \tilde{c}_{23} \tilde{s}_{13}\simeq
\sqrt{{2}/{5}}\sqrt{{M_\mu}/{M_\tau}}\approx 8.9^\circ$.
Thus, in our model, the small value of $\theta_{13}$  
is determined by an order unity term times
the small charged lepton ratio $\sqrt{M_\mu/M_\tau}$.

\begin{table}[tdp]
\captionsetup{justification=raggedright, singlelinecheck=false}
\caption{Comparison between our results (with and without symmetry)
and the best fit given by pdgLive\cite{pdg2013}. 
Note that the octant of of $\theta_{23}$ is not presently known;
our values prefer $\theta_{23}>\pi/4$. With symmetry, $c_\nu $ gets fixed,
leading to 
$\theta_{23}\simeq 51.6^\circ$.}
\begin{center}
\vskip-12pt
\begin{tabular}{|c|c|c|c|}
\hline
 & $\sin^2(2\theta_{12})$ &$\sin^2(2\theta_{23})$ & $\sin^2(2\theta_{13})$ \\
\hline
PDG best fit & $0.857{+0.023\atop -0.025}$ & $>0.95\ (90\% {\rm CL})$, 
&  $0.095\pm 0.010$ \\
 & ($1\sigma$ error) & 
 $(36^\circ \lsim \theta_{23} \lsim 54^\circ)$ & ($1\sigma$ error) \\
\hline 
Our values   &0.860 &$\theta_{23}\simeq 51.5^\circ$ &0.093 \\ \cline{2-4}
w/o symmetry   & 0.910   & $\theta_{23}\simeq 38.5^\circ$ &  0.143  \\ \hline
$c_\nu=0.2508$ &0.860 & 0.95&  0.093  \\ \hline
\end{tabular}
\end{center}
\label{table:one}
\end{table}%

{\sl Discussion:}
Before imposing the symmetry, we assume:
(i) The charged leptons and the neutrinos each have different mass 
and Weak eigenstates.
(ii) Mass matrices are hermitian and arranged in specific forms.
(iii) $\mathbf{M}_\ell$ has $M_{11}=M_{22}=M_\mu$. 
(iv) $\mathbf{m}$ 
has equal diagonal elements, which forces $\tilde\theta_{12}$ to be maximal. 

At this point there are two remaining 
charged-lepton parameters, which can be adjusted to yield 
the three physical charged-lepton masses.
The three parameters of the neutrino mass matrix are fixed to accommodate 
$\delta m^2_{21}$, $|\delta m^2_{32}|$, and $\theta_{23}$.
Then the remaining mixing angles are correctly reproduced.
One finds that only the inverted mass hierarchy is possible.

The {\sl ad hoc} assumptions (ii)-(iv) on the mass matrices can be replaced by 
the choice of the symmetry, which also fixes, otherwise undetermined,
the value of $\tilde{\theta}_{23}$, 
demonstrating the advantage of the symmetry argument.
The symmetry imposed naturally favors Majorana neutrinos,
because the Dirac neutrino masses become degenerate,
a result in conflict with observational inference. 

The symmetries we have imposed may appear to have rather arbitrary 
symmetry parameters, $c_\nu$ for Dirac and $s_{\rm M}$ for Majorana.
However, being emergent symmetries, they are not unusual. The main point is
the existence of such unbroken emergent symmetries that relate
all other parameters to lead to good agreement with all neutrino-mixing 
phenomenology. 
From the high energy point of view, these symmetries are 
in some sense ``hidden" because they do not emerge until the symmetry violating
terms in the EW lagrangian are gauged away due to the spontaneous 
symmetry breaking.
%

Note that we keep all decimals in $c_\nu=0.2508$ to distinguish it from 
precisely 1/4, or $s_{\rm M}^2=0.5990$ from $3/5$. 
This is necessitated by the precision of the charged lepton mass measurements.
If we used $\tilde{t}_{13}=c_\nu =1/4$ or $s_{\rm M}^2=3/5$ to determine
the charged lepton mixing, and input measured values of $M_e$ and $M_\mu$,
we would get $M_\tau\simeq 1788$ MeV. This result for $M_\tau$ is within $1\%$
of the measured value $M_\tau\simeq 1776.82\pm 0.16$ MeV, but still
$70\sigma$ away.  

We conclude with mention of another interesting 
feature\cite{La:2013gga} of our model.
The largest ratio in the charged lepton mass matrix, eq.(\ref{Mlep}), 
is $M_{33}/M_\mu\simeq 16$, and yet we obtain the huge charged 
lepton mass hierarchy $M_\tau/M_e \sim 4\times 10^3$. 
Furthermore, this is linked to the smallness of $s_{13}$ 
in terms of the smallness of $\tilde{s}_{13}\simeq \sqrt{M_\mu/M_\tau}$.
So the model we have presented could be a clue to deeper
understanding of lepton flavor physics.

\smallskip
\noindent
{\bf Acknowledgements}:
We thank Tom Kephart for conversations on the discrete groups.
HL also thanks David Ernst for his interests in this approach.
This research was supported in part by Department of Energy 
grant DE-FG05-85ER40226, and by Vanderbilt University.

\end{document}